\documentstyle[dina4,epsf]{article}

\begin{document}

\title{The Quark Structure of Light Mesons}

\author{B.\,C.\,Metsch,
        H.\,R.\,Petry\\ 
         Institut f\"ur Theoretische Kernphysik,\\
         Universit\"at Bonn, Nu{\ss}allee 14-16, 53115 Bonn, Germany\\
         {\footnotesize E-mail: metsch@itkp.uni-bonn.de}\\
}
\date{TK-96-25 \qquad 29.06.1996}

\maketitle

\section{Introduction} \label{I}
A glance at the experimental spectra of light mesons reveals
two general trends each with a conspicuous exception:
\begin{itemize}
     \item
          In general states that can be attributed to the same
          orbital angular momentum multiplets show only small
          spin-orbit splittings. Examples are $f_1(1285)$ and
          $f_2(1270)$, $a_3(2050)$ and $a_4(2040)$, $K_2(1770)$
          and $K_3^*(1780)$. Notable exceptions are the low
          positions of the $f_0(980)$ and $a_0(980)$.
     \item
          For every isovector state there is energetically
          degenerate an isoscalar partner. The best known example
          is of course $\rho(770)$ and $\omega(782)$, but also
          $h_1(1170)$ and $b_1(1235)$ and many other pairs,
          up to $a_6(2450)$ and $f_6(2510)$. This is of course
          nothing but the well known fact that the inter-quark
          forces are flavor symmetric. Note however, that this
          rule does {\em not} apply to the pseudoscalar mesons
          $\pi,\eta,\eta'$, which exhibit a mass splitting as large
          as any in the meson spectrum.
\end{itemize}

Any model of hadrons should address these observations. This seems to
be especially important for the identification of exotic mesonic
states, like hybrids, dimesonic states or the appreciation of the
predictions on glueballs from lattice-QCD~\cite{Bal93}. The most
successful framework for a coherent description of both meson and
baryon spectra certainly is the non-relativistic quark model. Here,
the basic assumptions are that the excitations of hadrons can be
effectively described by constituent quarks, that interact via
potentials in the framework of the the Schr\"odinger equation. A
particular version of the NRQM implementing confinement via a linear
potential in the spirit of the string model and substituting the
widely used Fermi-Breit interaction by an instanton induced effective
quark interaction, can simultaneously describe meson and baryon masses
up-to excitation energies of roughly 1 GeV~\cite{Bla90}. In this
framework the masses of the bulk of mesons are determined by the
confinement potential alone, thus avoiding unwanted spin-orbit
splittings. The instanton induced force selectively acts only on
pseudoscalar states and accounts for the large $\pi, \eta, \eta'$
splitting and mixing of isoscalars. However, it should be pointed out
that this treatment can be critizised at two points at least:
\begin{itemize}
\item Binding energies (especially of the ground state mesons)
  are too large compared to the constituent masses to justify a
  non-relativistic treatment;
\item The Schr\"odinger wave functions are incorrect at large
  energies and/or momentum transfers.
\end{itemize}
Although these seem to be rather formal objections they do matter
in practice. Let us consider the calculation of some electro-weak
decay observables for light mesons as gathered in
Table~\ref{decays}~\cite{Mue94}:

\begin{table}[htb] 
  \caption{
     Comparison of experimental and calculated electro-weak meson
     decay observables for the relativistic quark model 'RQM' and
     non-relativistic results 'NRQM'.
  }
 \label{decays}
\begin{center}
   \begin{tabular}{cccc}
  \hline
       Decay  &  exp.~\cite{PDG94} &  RQM  &  NRQM    \\
     \hline
          $f_{\pi}$ [MeV]
                  & 131.7$\pm$ 0.2   & 130  &  1440    \\
          $ f_K$ [MeV]
                  & 160.6 $\pm$ 1.4  &  180 &   730    \\
          $\Gamma(\pi^0 \rightarrow \gamma\gamma)$ [eV]
                  & 7.8 $\pm$ 0.5      & 7.6  & 30000    \\
          $\Gamma(\eta \rightarrow \gamma\gamma)$ [eV]
                  & 460  $\pm$ 5       & 440  & 18500    \\
          $\Gamma(\eta' \rightarrow \gamma\gamma)$ [eV]
                  & 4510 $\pm$ 260     & 2900 &   750    \\
          $\Gamma(\rho \rightarrow e^+e^-)$ [keV]
                  & 6.8 $\pm$ 0.3     & 6.8  &  8.95    \\
          $\Gamma(\omega \rightarrow e^+e^-)$ [keV]
                  & 0.60 $\pm$ 0.02   & 0.73 &  0.96    \\
          $\Gamma(\phi \rightarrow e^+e^-)$ [keV]
                  & 1.37 $\pm$ 0.05  & 1.24 &  2.06    \\
\hline
\end{tabular}
\end{center}
\end{table}

\begin{figure}[htb]\label{strange}
  \begin{center}
\begin{picture}(0,0)%
\epsfbox{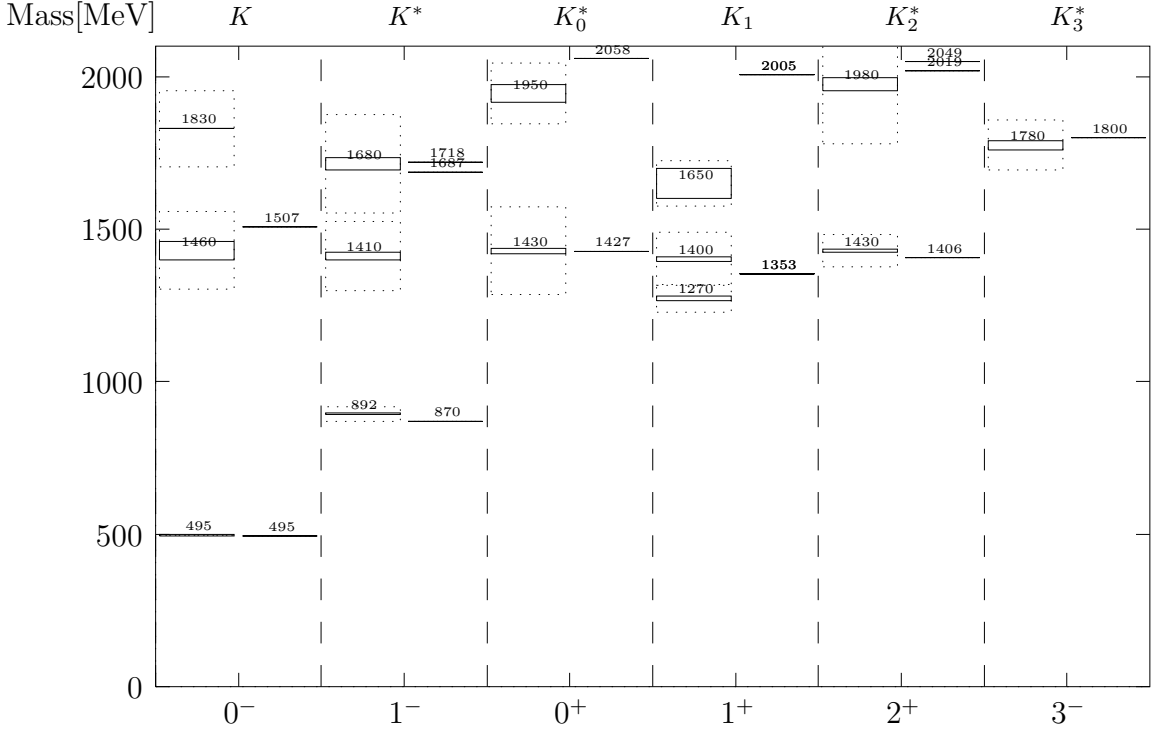}%
\end{picture}%
\setlength{\unitlength}{0.004033in}%
\begingroup\makeatletter\ifx\SetFigFont\undefined
\def\x#1#2#3#4#5#6#7\relax{\def\x{#1#2#3#4#5#6}}%
\expandafter\x\fmtname xxxxxx\relax \def\y{splain}%
\ifx\x\y   
\gdef\SetFigFont#1#2#3{%
  \ifnum #1<17\tiny\else \ifnum #1<20\small\else
  \ifnum #1<24\normalsize\else \ifnum #1<29\large\else
  \ifnum #1<34\Large\else \ifnum #1<41\LARGE\else
     \huge\fi\fi\fi\fi\fi\fi
  \csname #3\endcsname}%
\else
\gdef\SetFigFont#1#2#3{\begingroup
  \count@#1\relax \ifnum 25<\count@\count@25\fi
  \def\x{\endgroup\@setsize\SetFigFont{#2pt}}%
  \expandafter\x
    \csname \romannumeral\the\count@ pt\expandafter\endcsname
    \csname @\romannumeral\the\count@ pt\endcsname
  \csname #3\endcsname}%
\fi
\fi\endgroup
\begin{picture}(1311,941)(209,-132)
\put(209,-99){\makebox(0,0)[rb]{\smash{\SetFigFont{12}{14.4}{rm}0}}}
\put(209, 99){\makebox(0,0)[rb]{\smash{\SetFigFont{12}{14.4}{rm}500}}}
\put(209,298){\makebox(0,0)[rb]{\smash{\SetFigFont{12}{14.4}{rm}1000}}}
\put(209,496){\makebox(0,0)[rb]{\smash{\SetFigFont{12}{14.4}{rm}1500}}}
\put(209,694){\makebox(0,0)[rb]{\smash{\SetFigFont{12}{14.4}{rm}2000}}}
\put(228,-132){\makebox(0,0)[b]{\smash{\SetFigFont{12}{14.4}{rm} }}}
\put(336,-132){\makebox(0,0)[b]{\smash{\SetFigFont{12}{14.4}{rm}$0^{-}$}}}
\put(551,-132){\makebox(0,0)[b]{\smash{\SetFigFont{12}{14.4}{rm}$1^{-}$}}}
\put(766,-132){\makebox(0,0)[b]{\smash{\SetFigFont{12}{14.4}{rm}$0^{+}$}}}
\put(982,-132){\makebox(0,0)[b]{\smash{\SetFigFont{12}{14.4}{rm}$1^{+}$}}}
\put(1197,-132){\makebox(0,0)[b]{\smash{\SetFigFont{12}{14.4}{rm}$2^{+}$}}}
\put(1412,-132){\makebox(0,0)[b]{\smash{\SetFigFont{12}{14.4}{rm}$3^{-}$}}}
\put(336,776){\makebox(0,0)[b]{\smash{\SetFigFont{12}{14.4}{
rm}{\normalsize $K$}}}}
\put(551,776){\makebox(0,0)[b]{\smash{\SetFigFont{12}{14.4}{
rm}{\normalsize $K^*$}}}}
\put(766,776){\makebox(0,0)[b]{\smash{\SetFigFont{12}{14.4}{
rm}{\normalsize $K_0^*$}}}}
\put(982,776){\makebox(0,0)[b]{\smash{\SetFigFont{12}{14.4}{
rm}{\normalsize $K_1$}}}}
\put(1197,776){\makebox(0,0)[b]{\smash{\SetFigFont{12}{14.4}{
rm}{\normalsize $K_2^*$}}}}
\put(1412,776){\makebox(0,0)[b]{\smash{\SetFigFont{12}{14.4}{
rm}{\normalsize $K_3^*$}}}}
\put(282,119){\makebox(0,0)[b]{\smash{\SetFigFont{12}{14.4}{rm}{\tiny   495}}}}
\put(282,489){\makebox(0,0)[b]{\smash{\SetFigFont{12}{14.4}{rm}{\tiny  1460}}}}
\put(282,648){\makebox(0,0)[b]{\smash{\SetFigFont{12}{14.4}{rm}{\tiny  1830}}}}
\put(390,118){\makebox(0,0)[b]{\smash{\SetFigFont{12}{14.4}{rm}{\tiny   495}}}}
\put(390,520){\makebox(0,0)[b]{\smash{\SetFigFont{12}{14.4}{rm}{\tiny  1507}}}}
\put(497,276){\makebox(0,0)[b]{\smash{\SetFigFont{12}{14.4}{rm}{\tiny   892}}}}
\put(497,482){\makebox(0,0)[b]{\smash{\SetFigFont{12}{14.4}{rm}{\tiny  1410}}}}
\put(497,602){\makebox(0,0)[b]{\smash{\SetFigFont{12}{14.4}{rm}{\tiny  1680}}}}
\put(605,267){\makebox(0,0)[b]{\smash{\SetFigFont{12}{14.4}{rm}{\tiny   870}}}}
\put(605,591){\makebox(0,0)[b]{\smash{\SetFigFont{12}{14.4}{rm}{\tiny  1687}}}}
\put(605,604){\makebox(0,0)[b]{\smash{\SetFigFont{12}{14.4}{rm}{\tiny  1718}}}}
\put(713,489){\makebox(0,0)[b]{\smash{\SetFigFont{12}{14.4}{rm}{\tiny  1430}}}}
\put(713,693){\makebox(0,0)[b]{\smash{\SetFigFont{12}{14.4}{rm}{\tiny  1950}}}}
\put(820,488){\makebox(0,0)[b]{\smash{\SetFigFont{12}{14.4}{rm}{\tiny  1427}}}}
\put(820,738){\makebox(0,0)[b]{\smash{\SetFigFont{12}{14.4}{rm}{\tiny  2058}}}}
\put(928,427){\makebox(0,0)[b]{\smash{\SetFigFont{12}{14.4}{rm}{\tiny  1270}}}}
\put(928,478){\makebox(0,0)[b]{\smash{\SetFigFont{12}{14.4}{rm}{\tiny 1400}}}}
\put(928,576){\makebox(0,0)[b]{\smash{\SetFigFont{12}{14.4}{rm}{\tiny 1650}}}}
\put(1036,459){\makebox(0,0)[b]{\smash{\SetFigFont{12}{14.4}{
rm}{\tiny 1353}}}}
\put(1036,459){\makebox(0,0)[b]{\smash{\SetFigFont{12}{14.4}{rm}{\tiny 1353}}}}
\put(1036,718){\makebox(0,0)[b]{\smash{\SetFigFont{12}{14.4}{rm}{\tiny 2005}}}}
\put(1036,718){\makebox(0,0)[b]{\smash{\SetFigFont{12}{14.4}{rm}{\tiny 2005}}}}
\put(1143,489){\makebox(0,0)[b]{\smash{\SetFigFont{12}{14.4}{rm}{\tiny 1430}}}}
\put(1143,705){\makebox(0,0)[b]{\smash{\SetFigFont{12}{14.4}{rm}{\tiny 1980}}}}
\put(1251,480){\makebox(0,0)[b]{\smash{\SetFigFont{12}{14.4}{rm}{\tiny 1406}}}}
\put(1251,723){\makebox(0,0)[b]{\smash{\SetFigFont{12}{14.4}{rm}{\tiny 2019}}}}
\put(1251,735){\makebox(0,0)[b]{\smash{\SetFigFont{12}{14.4}{rm}{\tiny 2049}}}}
\put(1359,626){\makebox(0,0)[b]{\smash{\SetFigFont{12}{14.4}{rm}{\tiny 1780}}}}
\put(1466,636){\makebox(0,0)[b]{\smash{\SetFigFont{12}{14.4}{rm}{\tiny 1800}}}}
\put(228,776){\makebox(0,0)[rb]{\smash{\SetFigFont{12}{14.4}{rm}Mass[MeV]}}}
\end{picture}
\vspace{1.0ex}
  \caption{ 
    Strange meson spectrum. In the left part of each column the
    experimental resonance position {\protect \cite{PDG94}} and its
    error is indicated by a rectangle, the total decay width is
    given by a dotted rectangle; the lines in the right part
    of each column represent the calculated masses.
  }
  \end{center}
\end{figure}

\begin{figure}[hbt] \label{isoscalar}
  \begin{center}
\begin{picture}(0,0)%
\epsfbox{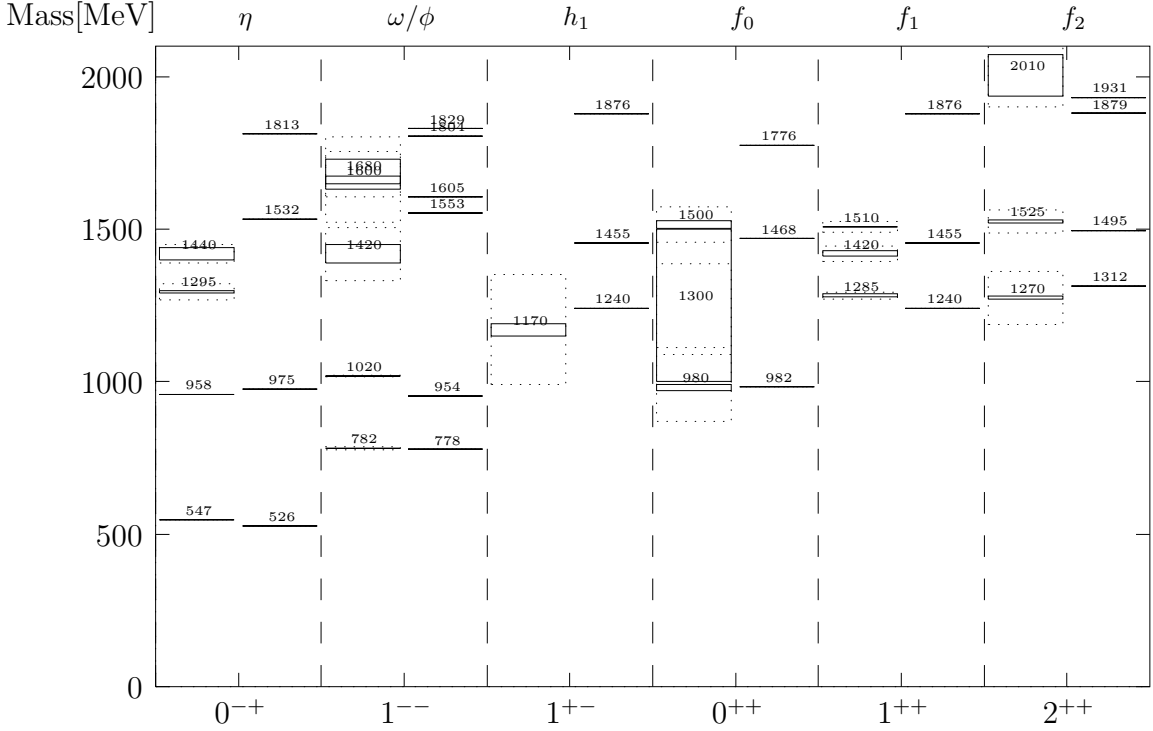}%
\end{picture}%
\setlength{\unitlength}{0.004033in}%
\begingroup\makeatletter\ifx\SetFigFont\undefined
\def\x#1#2#3#4#5#6#7\relax{\def\x{#1#2#3#4#5#6}}%
\expandafter\x\fmtname xxxxxx\relax \def\y{splain}%
\ifx\x\y   
\gdef\SetFigFont#1#2#3{%
  \ifnum #1<17\tiny\else \ifnum #1<20\small\else
  \ifnum #1<24\normalsize\else \ifnum #1<29\large\else
  \ifnum #1<34\Large\else \ifnum #1<41\LARGE\else
     \huge\fi\fi\fi\fi\fi\fi
  \csname #3\endcsname}%
\else
\gdef\SetFigFont#1#2#3{\begingroup
  \count@#1\relax \ifnum 25<\count@\count@25\fi
  \def\x{\endgroup\@setsize\SetFigFont{#2pt}}%
  \expandafter\x
    \csname \romannumeral\the\count@ pt\expandafter\endcsname
    \csname @\romannumeral\the\count@ pt\endcsname
  \csname #3\endcsname}%
\fi
\fi\endgroup
\begin{picture}(1311,941)(209,-132)
\put(209,-99){\makebox(0,0)[rb]{\smash{\SetFigFont{12}{14.4}{rm}0}}}
\put(209, 99){\makebox(0,0)[rb]{\smash{\SetFigFont{12}{14.4}{rm}500}}}
\put(209,298){\makebox(0,0)[rb]{\smash{\SetFigFont{12}{14.4}{rm}1000}}}
\put(209,496){\makebox(0,0)[rb]{\smash{\SetFigFont{12}{14.4}{rm}1500}}}
\put(209,694){\makebox(0,0)[rb]{\smash{\SetFigFont{12}{14.4}{rm}2000}}}
\put(228,-132){\makebox(0,0)[b]{\smash{\SetFigFont{12}{14.4}{rm} }}}
\put(336,-132){\makebox(0,0)[b]{\smash{\SetFigFont{12}{14.4}{rm}$0^{-+}$}}}
\put(551,-132){\makebox(0,0)[b]{\smash{\SetFigFont{12}{14.4}{rm}$1^{--}$}}}
\put(766,-132){\makebox(0,0)[b]{\smash{\SetFigFont{12}{14.4}{rm}$1^{+-}$}}}
\put(982,-132){\makebox(0,0)[b]{\smash{\SetFigFont{12}{14.4}{rm}$0^{++}$}}}
\put(1197,-132){\makebox(0,0)[b]{\smash{\SetFigFont{12}{14.4}{rm}$1^{++}$}}}
\put(1412,-132){\makebox(0,0)[b]{\smash{\SetFigFont{12}{14.4}{rm}$2^{++}$}}}
\put(336,776){\makebox(0,0)[b]{\smash{\SetFigFont{12}{14.4}{rm}{
\normalsize $\eta$}}}}
\put(551,776){\makebox(0,0)[b]{\smash{\SetFigFont{12}{14.4}{rm}{
\normalsize $\omega/\phi$}}}}
\put(766,776){\makebox(0,0)[b]{\smash{\SetFigFont{12}{14.4}{rm}{
\normalsize $h_1$}}}}
\put(982,776){\makebox(0,0)[b]{\smash{\SetFigFont{12}{14.4}{rm}{
\normalsize $f_0$}}}}
\put(1197,776){\makebox(0,0)[b]{\smash{\SetFigFont{12}{14.4}{rm}{
\normalsize $f_1$}}}}
\put(1412,776){\makebox(0,0)[b]{\smash{\SetFigFont{12}{14.4}{rm}{
\normalsize $f_2$}}}}
\put(282,139){\makebox(0,0)[b]{\smash{\SetFigFont{12}{14.4}{rm}{\tiny  547}}}}
\put(282,302){\makebox(0,0)[b]{\smash{\SetFigFont{12}{14.4}{rm}{\tiny  958}}}}
\put(282,436){\makebox(0,0)[b]{\smash{\SetFigFont{12}{14.4}{rm}{\tiny 1295}}}}
\put(282,485){\makebox(0,0)[b]{\smash{\SetFigFont{12}{14.4}{rm}{\tiny 1440}}}}
\put(390,131){\makebox(0,0)[b]{\smash{\SetFigFont{12}{14.4}{rm}{\tiny  526}}}}
\put(390,309){\makebox(0,0)[b]{\smash{\SetFigFont{12}{14.4}{rm}{\tiny  975}}}}
\put(390,530){\makebox(0,0)[b]{\smash{\SetFigFont{12}{14.4}{rm}{\tiny 1532}}}}
\put(390,641){\makebox(0,0)[b]{\smash{\SetFigFont{12}{14.4}{rm}{\tiny 1813}}}}
\put(497,232){\makebox(0,0)[b]{\smash{\SetFigFont{12}{14.4}{rm}{\tiny  782}}}}
\put(497,326){\makebox(0,0)[b]{\smash{\SetFigFont{12}{14.4}{rm}{\tiny 1020}}}}
\put(497,485){\makebox(0,0)[b]{\smash{\SetFigFont{12}{14.4}{rm}{\tiny 1420}}}}
\put(497,581){\makebox(0,0)[b]{\smash{\SetFigFont{12}{14.4}{rm}{\tiny 1600}}}}
\put(497,588){\makebox(0,0)[b]{\smash{\SetFigFont{12}{14.4}{rm}{\tiny 1680}}}}
\put(605,230){\makebox(0,0)[b]{\smash{\SetFigFont{12}{14.4}{rm}{\tiny  778}}}}
\put(605,300){\makebox(0,0)[b]{\smash{\SetFigFont{12}{14.4}{rm}{\tiny  954}}}}
\put(605,538){\makebox(0,0)[b]{\smash{\SetFigFont{12}{14.4}{rm}{\tiny 1553}}}}
\put(605,559){\makebox(0,0)[b]{\smash{\SetFigFont{12}{14.4}{rm}{\tiny 1605}}}}
\put(605,638){\makebox(0,0)[b]{\smash{\SetFigFont{12}{14.4}{rm}{\tiny 1804}}}}
\put(605,647){\makebox(0,0)[b]{\smash{\SetFigFont{12}{14.4}{rm}{\tiny 1829}}}}
\put(713,386){\makebox(0,0)[b]{\smash{\SetFigFont{12}{14.4}{rm}{\tiny 1170}}}}
\put(820,414){\makebox(0,0)[b]{\smash{\SetFigFont{12}{14.4}{rm}{\tiny 1240}}}}
\put(820,499){\makebox(0,0)[b]{\smash{\SetFigFont{12}{14.4}{rm}{\tiny 1455}}}}
\put(820,666){\makebox(0,0)[b]{\smash{\SetFigFont{12}{14.4}{rm}{\tiny 1876}}}}
\put(928,311){\makebox(0,0)[b]{\smash{\SetFigFont{12}{14.4}{rm}{\tiny  980}}}}
\put(928,418){\makebox(0,0)[b]{\smash{\SetFigFont{12}{14.4}{rm}{\tiny 1300}}}}
\put(928,523){\makebox(0,0)[b]{\smash{\SetFigFont{12}{14.4}{rm}{\tiny 1500}}}}
\put(1036,311){\makebox(0,0)[b]{\smash{\SetFigFont{12}{14.4}{rm}{\tiny  982}}}}
\put(1036,504){\makebox(0,0)[b]{\smash{\SetFigFont{12}{14.4}{rm}{\tiny 1468}}}}
\put(1036,626){\makebox(0,0)[b]{\smash{\SetFigFont{12}{14.4}{rm}{\tiny 1776}}}}
\put(1143,430){\makebox(0,0)[b]{\smash{\SetFigFont{12}{14.4}{rm}{\tiny 1285}}}}
\put(1143,485){\makebox(0,0)[b]{\smash{\SetFigFont{12}{14.4}{rm}{\tiny 1420}}}}
\put(1143,520){\makebox(0,0)[b]{\smash{\SetFigFont{12}{14.4}{rm}{\tiny 1510}}}}
\put(1251,414){\makebox(0,0)[b]{\smash{\SetFigFont{12}{14.4}{rm}{\tiny 1240}}}}
\put(1251,499){\makebox(0,0)[b]{\smash{\SetFigFont{12}{14.4}{rm}{\tiny 1455}}}}
\put(1251,666){\makebox(0,0)[b]{\smash{\SetFigFont{12}{14.4}{rm}{\tiny 1876}}}}
\put(1359,428){\makebox(0,0)[b]{\smash{\SetFigFont{12}{14.4}{rm}{\tiny 1270}}}}
\put(1359,527){\makebox(0,0)[b]{\smash{\SetFigFont{12}{14.4}{rm}{\tiny 1525}}}}
\put(1359,717){\makebox(0,0)[b]{\smash{\SetFigFont{12}{14.4}{rm}{\tiny 2010}}}}
\put(1466,443){\makebox(0,0)[b]{\smash{\SetFigFont{12}{14.4}{rm}{\tiny 1312}}}}
\put(1466,515){\makebox(0,0)[b]{\smash{\SetFigFont{12}{14.4}{rm}{\tiny 1495}}}}
\put(1466,667){\makebox(0,0)[b]{\smash{\SetFigFont{12}{14.4}{rm}{\tiny 1879}}}}
\put(1466,688){\makebox(0,0)[b]{\smash{\SetFigFont{12}{14.4}{rm}{\tiny 1931}}}}
\put(228,776){\makebox(0,0)[rb]{\smash{\SetFigFont{12}{14.4}{rm}Mass[MeV]}}}
\end{picture}
\vspace{1ex}
  \caption{
    Isoscalar meson spectrum. See also caption to Fig.\protect\ref{strange}.
  }
  \end{center}
\end{figure}
  
\begin{figure}[htb]\label{isovector}
  \begin{center}
\begin{picture}(0,0)%
\epsfbox{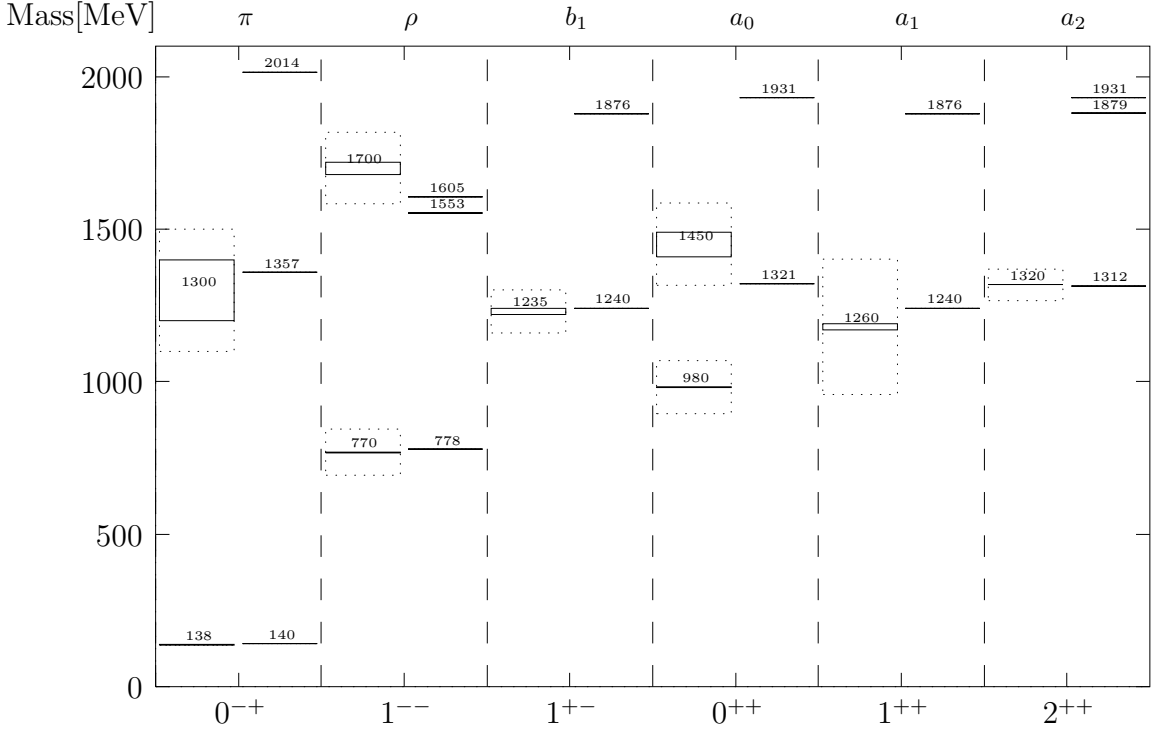}%
\end{picture}%
\setlength{\unitlength}{0.004033in}%
\begingroup\makeatletter\ifx\SetFigFont\undefined
\def\x#1#2#3#4#5#6#7\relax{\def\x{#1#2#3#4#5#6}}%
\expandafter\x\fmtname xxxxxx\relax \def\y{splain}%
\ifx\x\y   
\gdef\SetFigFont#1#2#3{%
  \ifnum #1<17\tiny\else \ifnum #1<20\small\else
  \ifnum #1<24\normalsize\else \ifnum #1<29\large\else
  \ifnum #1<34\Large\else \ifnum #1<41\LARGE\else
     \huge\fi\fi\fi\fi\fi\fi
  \csname #3\endcsname}%
\else
\gdef\SetFigFont#1#2#3{\begingroup
  \count@#1\relax \ifnum 25<\count@\count@25\fi
  \def\x{\endgroup\@setsize\SetFigFont{#2pt}}%
  \expandafter\x
    \csname \romannumeral\the\count@ pt\expandafter\endcsname
    \csname @\romannumeral\the\count@ pt\endcsname
  \csname #3\endcsname}%
\fi
\fi\endgroup
\begin{picture}(1311,941)(209,-132)
\put(209,-99){\makebox(0,0)[rb]{\smash{\SetFigFont{12}{14.4}{rm}0}}}
\put(209, 99){\makebox(0,0)[rb]{\smash{\SetFigFont{12}{14.4}{rm}500}}}
\put(209,298){\makebox(0,0)[rb]{\smash{\SetFigFont{12}{14.4}{rm}1000}}}
\put(209,496){\makebox(0,0)[rb]{\smash{\SetFigFont{12}{14.4}{rm}1500}}}
\put(209,694){\makebox(0,0)[rb]{\smash{\SetFigFont{12}{14.4}{rm}2000}}}
\put(228,-132){\makebox(0,0)[b]{\smash{\SetFigFont{12}{14.4}{rm} }}}
\put(336,-132){\makebox(0,0)[b]{\smash{\SetFigFont{12}{14.4}{rm}$0^{-+}$}}}
\put(551,-132){\makebox(0,0)[b]{\smash{\SetFigFont{12}{14.4}{rm}$1^{--}$}}}
\put(766,-132){\makebox(0,0)[b]{\smash{\SetFigFont{12}{14.4}{rm}$1^{+-}$}}}
\put(982,-132){\makebox(0,0)[b]{\smash{\SetFigFont{12}{14.4}{rm}$0^{++}$}}}
\put(1197,-132){\makebox(0,0)[b]{\smash{\SetFigFont{12}{14.4}{rm}$1^{++}$}}}
\put(1412,-132){\makebox(0,0)[b]{\smash{\SetFigFont{12}{14.4}{rm}$2^{++}$}}}
\put(336,776){\makebox(0,0)[b]{\smash{\SetFigFont{12}{14.4}{rm}{
\normalsize $\pi$}}}}
\put(551,776){\makebox(0,0)[b]{\smash{\SetFigFont{12}{14.4}{rm}{
\normalsize $\rho$}}}}
\put(766,776){\makebox(0,0)[b]{\smash{\SetFigFont{12}{14.4}{rm}{
\normalsize $b_1$}}}}
\put(982,776){\makebox(0,0)[b]{\smash{\SetFigFont{12}{14.4}{rm}{
\normalsize $a_0$}}}}
\put(1197,776){\makebox(0,0)[b]{\smash{\SetFigFont{12}{14.4}{rm}{
\normalsize $a_1$}}}}
\put(1412,776){\makebox(0,0)[b]{\smash{\SetFigFont{12}{14.4}{rm}{
\normalsize $a_2$}}}}
\put(282,-24){\makebox(0,0)[b]{\smash{\SetFigFont{12}{14.4}{rm}{\tiny  138}}}}
\put(282,437){\makebox(0,0)[b]{\smash{\SetFigFont{12}{14.4}{rm}{\tiny 1300}}}}
\put(390,-23){\makebox(0,0)[b]{\smash{\SetFigFont{12}{14.4}{rm}{\tiny  140}}}}
\put(390,460){\makebox(0,0)[b]{\smash{\SetFigFont{12}{14.4}{rm}{\tiny 1357}}}}
\put(390,721){\makebox(0,0)[b]{\smash{\SetFigFont{12}{14.4}{rm}{\tiny 2014}}}}
\put(497,227){\makebox(0,0)[b]{\smash{\SetFigFont{12}{14.4}{rm}{\tiny  770}}}}
\put(497,596){\makebox(0,0)[b]{\smash{\SetFigFont{12}{14.4}{rm}{\tiny 1700}}}}
\put(605,230){\makebox(0,0)[b]{\smash{\SetFigFont{12}{14.4}{rm}{\tiny  778}}}}
\put(605,538){\makebox(0,0)[b]{\smash{\SetFigFont{12}{14.4}{rm}{\tiny 1553}}}}
\put(605,559){\makebox(0,0)[b]{\smash{\SetFigFont{12}{14.4}{rm}{\tiny 1605}}}}
\put(713,410){\makebox(0,0)[b]{\smash{\SetFigFont{12}{14.4}{rm}{\tiny 1235}}}}
\put(820,414){\makebox(0,0)[b]{\smash{\SetFigFont{12}{14.4}{rm}{\tiny 1240}}}}
\put(820,666){\makebox(0,0)[b]{\smash{\SetFigFont{12}{14.4}{rm}{\tiny 1876}}}}
\put(928,312){\makebox(0,0)[b]{\smash{\SetFigFont{12}{14.4}{rm}{\tiny  980}}}}
\put(928,497){\makebox(0,0)[b]{\smash{\SetFigFont{12}{14.4}{rm}{\tiny 1450}}}}
\put(1036,446){\makebox(0,0)[b]{\smash{\SetFigFont{12}{14.4}{rm}{\tiny 1321}}}}
\put(1036,688){\makebox(0,0)[b]{\smash{\SetFigFont{12}{14.4}{rm}{\tiny 1931}}}}
\put(1143,390){\makebox(0,0)[b]{\smash{\SetFigFont{12}{14.4}{rm}{\tiny 1260}}}}
\put(1251,414){\makebox(0,0)[b]{\smash{\SetFigFont{12}{14.4}{rm}{\tiny 1240}}}}
\put(1251,666){\makebox(0,0)[b]{\smash{\SetFigFont{12}{14.4}{rm}{\tiny 1876}}}}
\put(1359,445){\makebox(0,0)[b]{\smash{\SetFigFont{12}{14.4}{rm}{\tiny 1320}}}}
\put(1466,443){\makebox(0,0)[b]{\smash{\SetFigFont{12}{14.4}{rm}{\tiny 1312}}}}
\put(1466,667){\makebox(0,0)[b]{\smash{\SetFigFont{12}{14.4}{rm}{\tiny 1879}}}}
\put(1466,688){\makebox(0,0)[b]{\smash{\SetFigFont{12}{14.4}{rm}{\tiny 1931}}}}
\put(228,776){\makebox(0,0)[rb]{\smash{\SetFigFont{12}{14.4}{rm}Mass[MeV]}}}
\end{picture}
\vspace{1.0ex}
  \caption{
    Isovector meson spectrum. See also caption to Fig.\protect\ref{strange}.
  }
  \end{center}
\end{figure}

Although the dilepton decays of the vector mesons can be more or less
accounted for non-relativistically, the discrepancies for the weak
decay constants amount up to an order of magnitude and the calculated
values for the $\gamma\gamma$-decay are beyond discussion. The
agreement with data can be drastically improved in a relativistically
covariant quark model we developed on the basis of the Bethe--Salpeter
equation, see section~\ref{II}. In section~\ref{III} we will present
some results for spectra, electro-weak
decays and form factors in this model. In section~\ref{IV} we will
demonstrate that in contrast to the non-relativistic model the
instanton induced interaction not only acts on pseudoscalar mesons but
also determines the structure and splittings of scalar mesons. Here we
will touch upon the consequences in particular for the possible
interpretation of the newly discovered $f_0(1500)$ resonance as a
glueball, see also~\cite{koch96}. The last section contains some
conclusions and an outlook.

\section{A relativistic quark model}\label{II}

In momentum space the Bethe--Salpeter equation for the amplitude
\begin{equation}\label{eq.1}
  \left[ \chi^{}_P(x) \right]_{\alpha\beta} =
  \left\langle\, 0 \,
     \left|\,
       T\,\Psi^1_{\alpha}(\frac{1}{2}x)\,
          \overline{\Psi}^2_{\beta}(-\frac{1}{2}x) \,
     \right|\, P \,
  \right\rangle ,
\end{equation}
reads~\cite{BS,Res94}:
\begin{eqnarray}\label{4}
  \chi^{}_P(p) &=& S^F_1(p_1)\, \int\! \frac{d^4 p'}{(2\pi)^4}\,
  [-i\,K(P,p,p')\,\chi^{}_P(p')]\, S^F_2(-p_2)
\end{eqnarray} 
where $p_1=\frac{1}{2}P+p ,\; p_2=\frac{1}{2}P-p$ denote the
momenta of the quark and antiquark respectively, $P$ is the four
momentum of the bound state and $S^F$ and $K$ are the Feynman
quark propagators and the irreducible quark interaction kernel.

We will construct the relativistic quark model very similar to
the non relativistic potential model. Therefore both $S_F$ and
$K$ are given by a phenomenological (but formally covariant) {\em
Ansatz} as follows:
\begin{itemize}
\item The propagators are assumed to be of the free form $S^F_i(p) =
  i/(p\!\!  /-m_i+i\epsilon)$, with an effective constituent-quark
  mass $m_i$;
\item It is assumed that the interaction kernel only depends on the
  components of $p$ and $p'$ perpendicular to $P$, i.e.
  $K(P,p,p') = V(p^{}_{_{\perp P}},p'_{_{\perp P}})$
  with $p^{}_{_{\perp P}} := p - (pP/P^2)P$.
\end{itemize}
Integrating in the bound state rest frame over the time component
and introducing the equal-time (or Salpeter) amplitude
\begin{equation}
  \Phi(\vec{p}\,) := \int\!
  dp^0\,\chi^{}_P(p^0,\vec{p}\,)\,\Big|_{P=(M,\vec{0}\,)}= \int\!
  dp^{}_{_{||P}}\,\chi^{}_P(p^{}_{_{||P}},p^{}_{_{\perp P}})\,
  \Big|_{P=(M,\vec{0}\,)}
\end{equation}
we thus arrive at the well-known Salpeter equation \cite{Sa}:
\begin{eqnarray} \label{SalEq}
     \Phi(\vec{p}\,) &=&
     \int \!\!\frac{d^3p'}{(2\pi)^3}\,
     \frac{\Lambda^-_1(\vec{p}\,)\,\gamma^0\,
     [(V(\vec{p},\vec{p}\,')\,\Phi(\vec{p}\,')]
     \,\gamma^0\,\Lambda^+_2(-\vec{p}\,)}
     {M+\omega_1+\omega_2}
      \nonumber \\
      &-&
     \int \!\!\frac{d^3p'}{(2\pi)^3}\,
     \frac{\Lambda^+_1(\vec{p}\,)\,\gamma^0\,
     [(V(\vec{p},\vec{p}\,')\,\Phi(\vec{p}\,')]
     \,\gamma^0\,\Lambda^-_2(-\vec{p}\,)}
     {M-\omega_1-\omega_2}
\end{eqnarray}
with the projectors $\Lambda^{\pm}_i = (\omega_i \pm
H_i)/(2\omega_i)$, the Dirac Hamiltonian
$H_i(\vec{p}\,)=\gamma^0(\vec{\gamma}\vec{p}+m_i)$ and
$\omega_i=(m^2_i+\vec{p\,}^2)^{1/2}$.

The amplitudes $\Phi$ have been calculated by solving the Salpeter
equation for a kernel including a confining interaction of the form 
\begin{equation}
 \int \! d^3p' \left[V_C^V(\vec{p},\vec{p}\,')\,
              \Phi(\vec{p}\,')\right] =
     -\int \! d^3p'\;{\cal
    V}((\vec{p}-\vec{p}\,')^2)\;
  \frac{1}{2} \left[
  \gamma^0\,\Phi(\vec{p}\,')\,\gamma^0
  + 1\,\Phi(\vec{p}\,')\,1 \right]
\label{conf}
\end{equation} 
where the spin structure was chosen such as to minimize spin
orbit effects. In coordinate space ${\cal V}$ is given by a
linearly rising potential ${\cal V}(|\vec x_q-\vec x_{\bar q}|) =
a+b|\vec x_q-\vec x_{\bar q}|$, in analogy to non-relativistic
quark models. We added a residual instanton-induced interaction
${\cal W}$ discovered by 't Hooft~\cite{Hoo76,SVZ80,Bla90,Mue94},
which acts only on pseudoscalar and scalar mesons and has the
form
\begin{eqnarray}\label{thokern}
 \int \! d^3p'\, [{\cal W}(\vec{p},\vec{p}\,')\,\Phi(\vec{p}\,')] = 
 4\,\,G \int \! d^3p'\,\,w_\lambda(\vec{p}-\vec{p}\,')\,\left[
  1\,\mbox{tr}\,\left(\Phi(\vec{p}\,')\right) +
  \gamma^5\,\mbox{tr}\,\left(\Phi(\vec{p}\,')\,\gamma^5\right)\,\right]
\end{eqnarray}
where $G(g,g')$ is a flavor matrix containing the coupling
constants $g$, $g'$. Here summation over flavor has been
suppressed and $w_\lambda$ is a regularizing Gaussian function
(see \cite{Mue94} for details).

To arrive at a covariant calculation of those transition matrix
elements with energy-momentum conservation for both particles,
the Bethe--Salpeter amplitude $\chi^{}_P(p)$ depending on the
relative four-momentum $p$ has to be known.  On the mass shell,
it can be reconstructed from the equal time amplitude
$\Phi(\vec{p}\,)$: From the Bethe--Salpeter equation the
meson-quark-anti\-quark vertex function $\Gamma_P(p) :=
[S^F_1(p_1)]^{-1}$ $\chi^{}_P(p)$ $[S^F_2(-p_2)]^{-1}$ is
computed in the rest frame from the equal-time amplitude as
\begin{eqnarray} 
  \Gamma_P(p^{}_{_{\perp P}})\left|_{_{P=(M,\vec{0}\,)}}\right.  \; =\;
\Gamma(\vec{p}\,) \; =\;
  -i\! \int\!\! \frac{d^3p'}{(2\pi)^4}
  \left[ V(\vec{p},\vec{p}\,')\Phi(\vec{p}\,')\right]
\label{vert}
\end{eqnarray} 
By a pure boost $\Lambda_P$ we then can calculate the BS
amplitude for any on-shell momentum $P$ of the bound-state by
\begin{equation}
  \chi^{}_P(p) = \;
  S_{\Lambda_P}^{}\;\;\chi^{}_{(M,\vec{0})}(\Lambda_P^{-1}p)\;\;
  S_{\Lambda_P}^{-1}.
\label{boo}
\end{equation}
The Salpeter equation (\ref{SalEq}) is solved numerically by expanding
the Salpeter amplitudes $\Phi$ in a finite basis and diagonalization
of the resulting RPA- type matrix~\cite{Res94}. Subsequently, the
vertex function (\ref{vert}) is reconstructed and used to calculate
elektroweak current matrix elements in the Mandelstam formalism, see
below and also~\cite{Mue94b}. The parameters of the model, i.e. the
constituent quark masses, $m_n$, $m_s$, confinement offset $a$ and
slope $b$ were adjusted to the Regge trajectories; the couplings $g$,
$g'$, and range $\lambda$ of the instanton induced force were chosen
to reproduce the pseudoscalar ground state masses.

\section{Spectra and decays of light mesons}\label{III}
The comparison of the experimental and calculated masses of
mesons on the Regge trajectories in Table 2 shows, that
the present {\em Ansatz} indeed successfully accounts for the
spectra at higher energies. The low energy part of the calculated
meson spectra are compared to experimental data in
Figs.1--3, for strange,
isoscalar and isovector particles, respectively. Note that with
the present interaction there are no large spin orbit effects,
which is in accordance with experiment for the strange mesons, but
certainly does not explain the low lying $a_0(980)$ resonance. We will
discuss this and the puzzling situation for the isoscalar, scalar
resonances in the following section.
\begin{table}[hbt] 
  \caption{
    Comparison of experimental \protect\cite{PDG94} and calculated
    masses (in MeV) of mesons on Regge trajectories.
  }
 \label{regge}
\begin{center}
   \begin{tabular}{ccc|ccc|ccc|ccc}
\hline
       \quad & exp. & calc. & \quad & exp. & calc. &
       \quad & exp. & calc. & \quad & exp. & calc. \\
     \hline
 $\omega$   &  782 &  778 & $\phi$   & 1020 &  954 &
 $\rho$     &  770 &  778 & $K^*$    &  892 &  870 \\  
 $f_2$      & 1270 & 1312 & $f_2'$   & 1525 & 1495 &
 $a_2$      & 1320 & 1312 & $K^*_2$  & 1430 & 1406 \\
 $\omega_3$ & 1670 & 1698 & $\phi_3$ & 1850 & 1900 &
 $\rho_3$   & 1690 & 1698 & $K^*_3$  & 1780 & 1800 \\
 $f_4$      & 2050 & 2011 & $f_4'$   & 2220 & 2230 &
 $a_4$      & 2040 & 2011 & $K^*_4$  & 2045 & 2121 \\
 $\omega_5$ &      & 2279 & $\phi_5$ &      & 2514 &
 $\rho_5$   & 2350 & 2279 & $K^*_5$  & 2380 & 2397 \\
 $f_6$      & 2510 & 2517 & $f_6'$   &      & 2766 &
 $a_6$      & 2450 & 2517 & $K^*_6$  &      & 2642 \\
\hline
\end{tabular}
\end{center}
\end{table}
The comparison of the experimental and calculated masses of
mesons on the Regge trajectories in Table~\ref{regge} shows, that
the present {\em Ansatz} indeed successfully accounts for the
spectra at higher energies. The low energy part of the calculated
meson spectra are compared to experimental data in
Figs.~\ref{strange},\ref{isovector},\ref{isoscalar}, for strange,
isovector and isoscalar particles, respectively. Note that with
the present interaction there are no large spin orbit effects,
which is in accordance with experiment for the strange mesons, but
certainly does not explain the low lying $a_0(980)$ resonance. We will
discuss this and the puzzling situation for the isoscalar, scalar
resonances in the following section.

\begin{figure}[htbp]
 \begin{minipage}[t]{6.5cm}
    \epsfxsize=6.5cm
    \epsffile{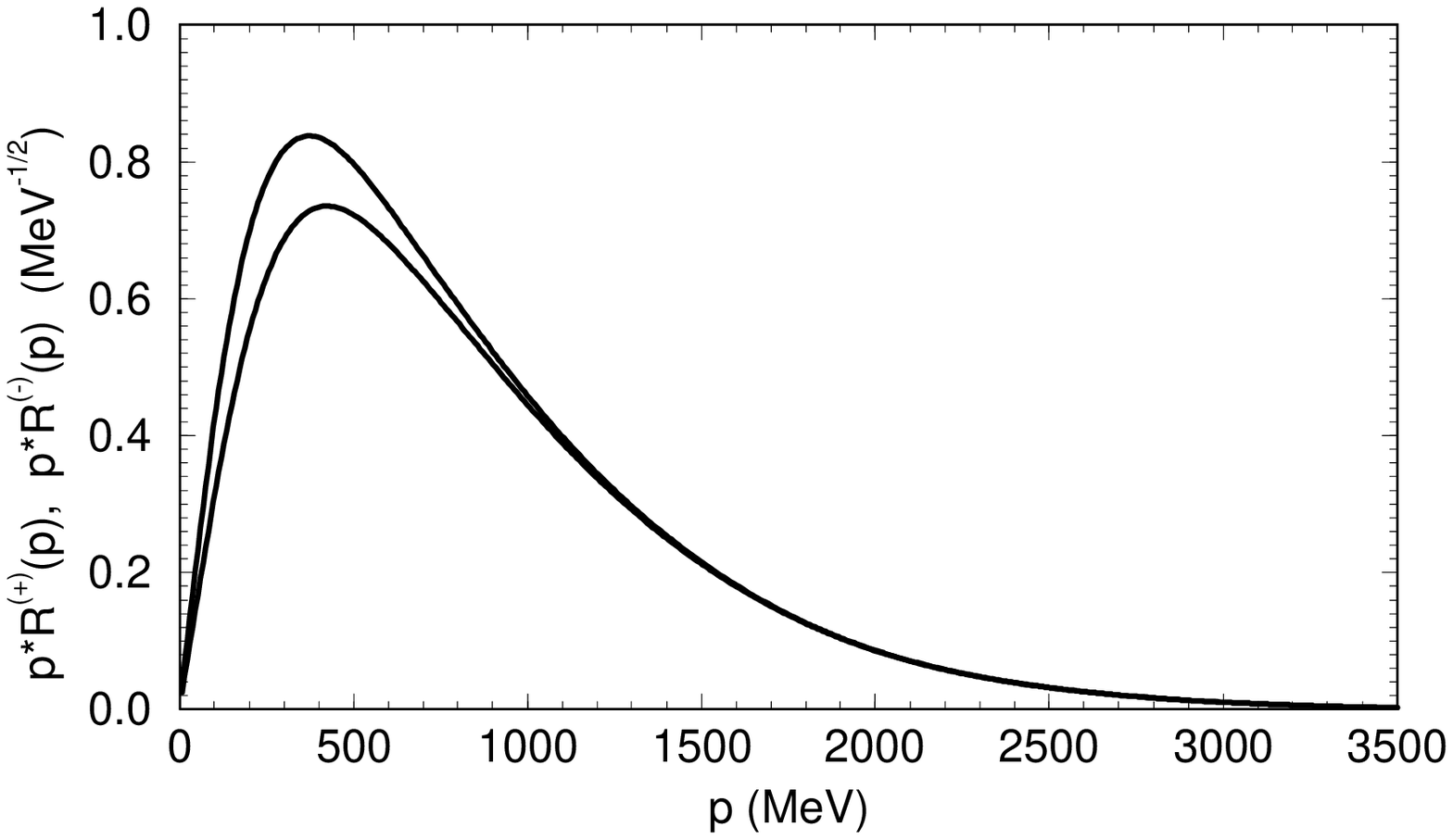}
     \vspace{0.5cm}
     \caption{Radial Pion amplitudes
     \(p\,{\cal R}^{(+)}_{00}(p)\) (upper component, upper curve) and
     \(p\,{\cal R}^{(-)}_{00}(p)\) (lower component, lower curve)
   }
   \label{pionamp}
\end{minipage}
\hfill
\begin{minipage}[t]{6.5cm}
    \epsfxsize=6.5cm
    \epsffile{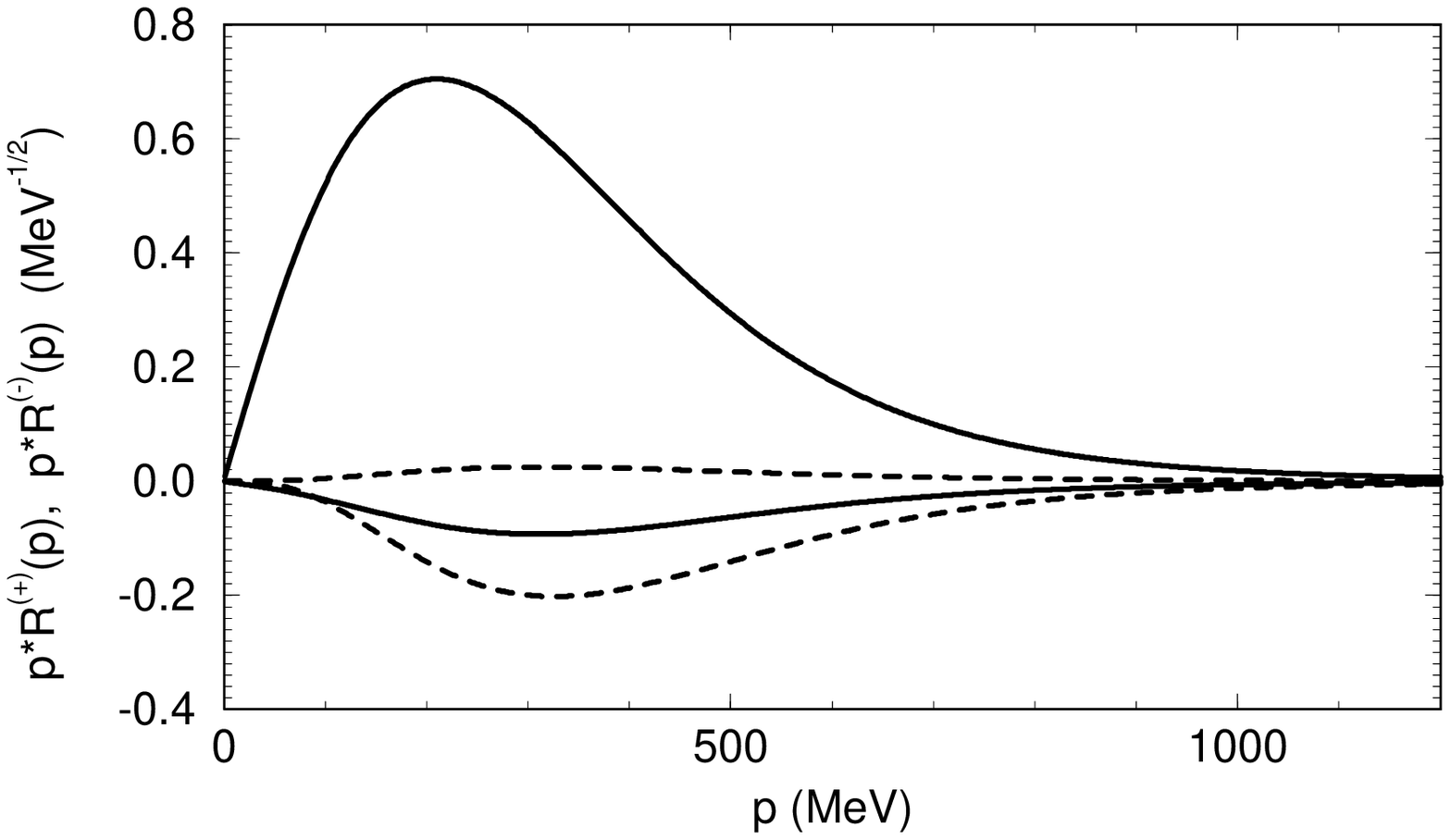}
 \vspace{0.5cm}
 \caption{Radial Rho amplitudes
   \(p\,{\cal R}^{(+)}_{01}(p)\) (upper s-wave component, upper solid curve),
   \(p\,{\cal R}^{(-)}_{01}(p)\) (lower s-wave component, lower solid curve),
   \(p\,{\cal R}^{(+)}_{21}(p)\) (upper d-wave component, upper dashed curve),
   \(p\,{\cal R}^{(-)}_{21}(p)\) (lower d-wave component, lower dashed curve)
   }
 \label{rhoamp}
\end{minipage}
\end{figure}

In order to appreciate the improvement as presented in
Table~\ref{decays} we show in Fig.~\ref{pionamp} and~\ref{rhoamp} the
radial part of the Salpeter amplitudes for a deeply bound state like
the pion and moderately bound state such as the $\rho$-meson,
respectively.  Whereas for the latter the relativistic components are
indeed small, and correspondingly the relativistic effects on
observables such as given in Table~\ref{decays} are moderate, for the
pion the upper and the lower component of the Salpeter amplitude are
of the same size. In the calculation of $f_\pi$ e.g. essentially their
difference enters and hence the correction by an order of magnitude.

The general prescription for calculating current matrix elements
between bound states has been given by Mandelstam~\cite{Man55},
The electromagnetic current operator may be calculated from the
BS amplitudes and a kernel \(K^{(\gamma)}\) that in its simplest
form corresponds to the impulse approximation and reads:
\begin{eqnarray}
  \lefteqn{K^{(\gamma)}_{\mu}(P,q,p,p') =} 
\label{kpertu}\\ & & -
  e_1\gamma_{\mu}^{(1)}\,{S^F_2}^{-1}(-P/2+p)\,\delta(p'-p+q/2)\; -
  e_2\gamma_{\mu}^{(2)}\,{S^F_1}^{-1}( P/2+p)\,\delta(p'-p-q/2)
  \nonumber
\end{eqnarray}
where \(p\) and \(p'\) denote the relative momenta of the
incoming and outgoing \(q\bar{q}\) pairs, $e_1$ and $-e_2$ are
the charges of the quark and antiquark, \(q=P-P'\) is the
momentum transfer of the photon.

For the electromagnetic current coupling e.g. to the first quark we
have explicitly
\begin{eqnarray}\label{current}
  \lefteqn{\left\langle\,P'\,\left|\,j_{\mu}^{(1)}(0)\,
     \right|\,P\,\right\rangle\; =\;
     - \; e_1\;\int\!\!\frac{d^4p}{(2\pi)^4}\;}\\
  & &     \mbox{tr}  \left\{
          \overline{\Gamma}_{P'}\left( (p-q/2)_{_{\perp P'}} \right)\;
          {S^F_1}(P/2+p-q)\;
          \gamma_{\mu}\;{S^F_1}(P/2+p)\;\Gamma^{}_{P}(p_{_{\perp P}})\;
     {S^F_2}(-P/2+p)\right\}\nonumber
\end{eqnarray} 
in terms of the vertex functions $\Gamma$, see Eq.(\ref{vert}),
given above. We like to point out, that this procedure respects
covariance and current conservation for the transitions studied.
However, in order to obtain a hermitian current we have to adopt
the additional prescription to take only the residue
contributions of the one-particle propagators in the expression
above, as otherwise the neglect of retardation effects would
yield a anti-hermitian principal-value integral, see
also~\cite{Mue94b}.

As an example for electromagnetic observables we compare the
calculated $\omega^0 \rightarrow \pi^0$ transition form factor in
the space-like region with an extrapolation of the experimental
data from the time-like region as measured in dilepton production.
It is seen that our results reproduce the vector dominance like
behavior of this form factor rather well however without
explicitly invoking this mechanism. For other examples of decay
widths and form factors, we refer to~\cite{Mue94b}.

\begin{figure}[htb]
  \centering
  \leavevmode
  \epsfxsize=0.6\textwidth
  \epsffile{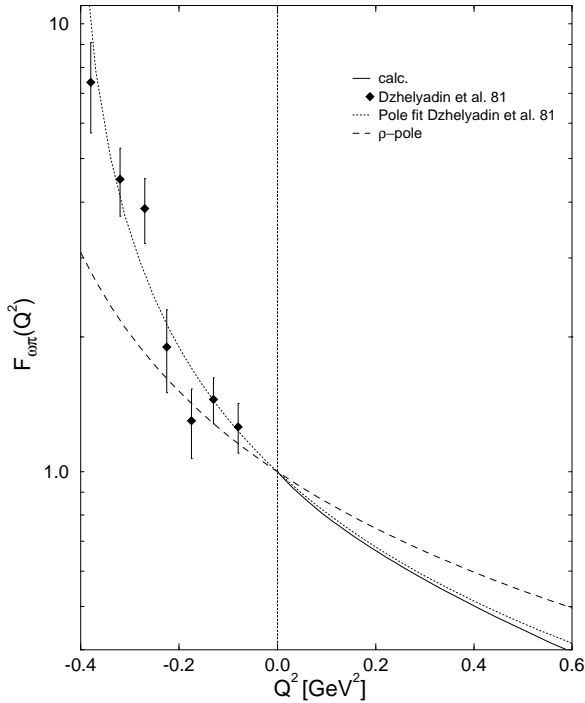}
  \caption{Comparison of the normalized \(\omega\pi\gamma^*\) form
    factor (solid line) in the space-like region with an extrapolation
    of experimental data in the time-like region
    \protect\cite{Dzh81} (dotted line) and with a $\rho$-pole {\em Ansatz}
    motivated by vector dominance (dashed line)}.
  \label{ffomegapi}
\end{figure}

\section{Instanton effects for pseudoscalar and scalar mesons}\label{IV}
\begin{figure}[htbp]
 \begin{minipage}[t]{6.5cm}
    \epsfxsize=6.5cm
  \epsffile{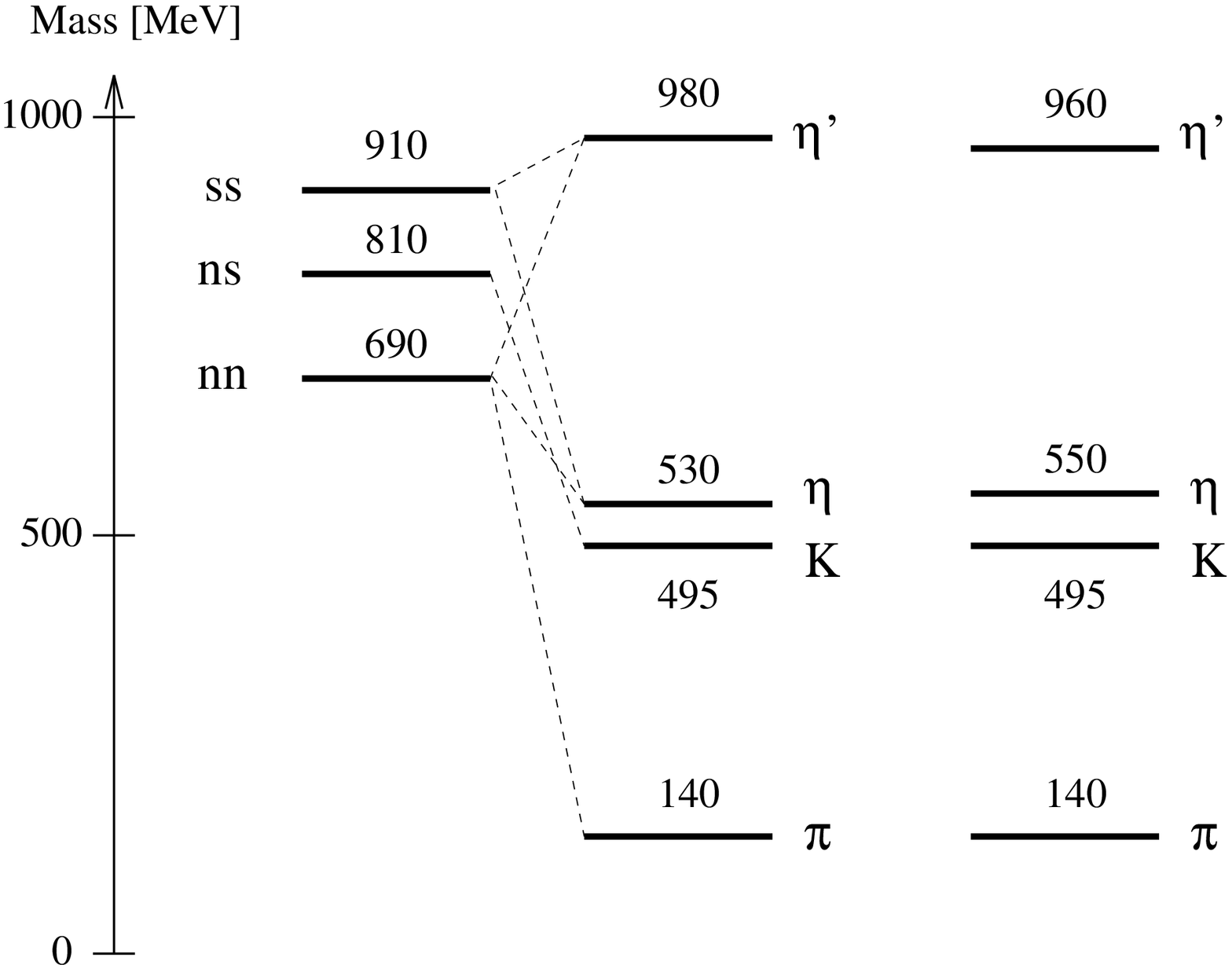}
\vspace{0.5cm}    
\caption{
    Schematic splitting of the pseudoscalar flavor nonet with confinement
    interaction (left), with confinement and instanton-induced force (middle)
    compared to the compilation by the Particle Data Group
    \protect{\cite{PDG94}} (right). 
}
  \label{pseudoscalars}
\end{minipage}
  \hfill
\begin{minipage}[t]{6.5cm}
    \epsfxsize=6.5cm
\epsffile{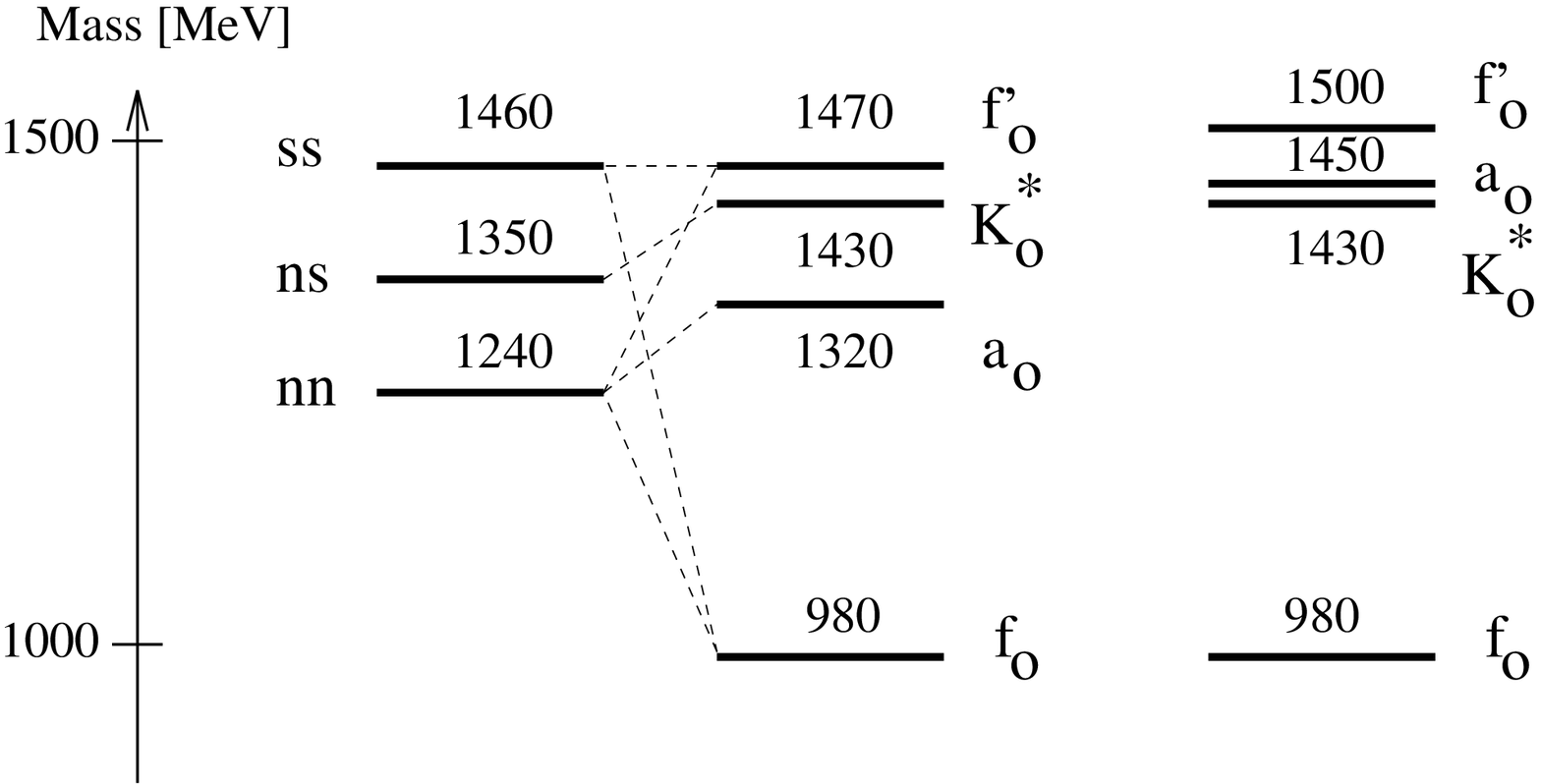}
\vspace{0.5cm}
\caption{
    Schematic splitting of the scalar flavor nonet with 
    confinement interaction (left), with confinement and instanton-induced
    force (middle) compared to the experimental
    spectrum interpreted as $q \bar q$ 
    states~\protect{\cite{PDG94,cb,amsler94}}
    (right).  
}
  \label{scalars}
  \end{minipage}
\end{figure}

\begin{figure}[htbp]
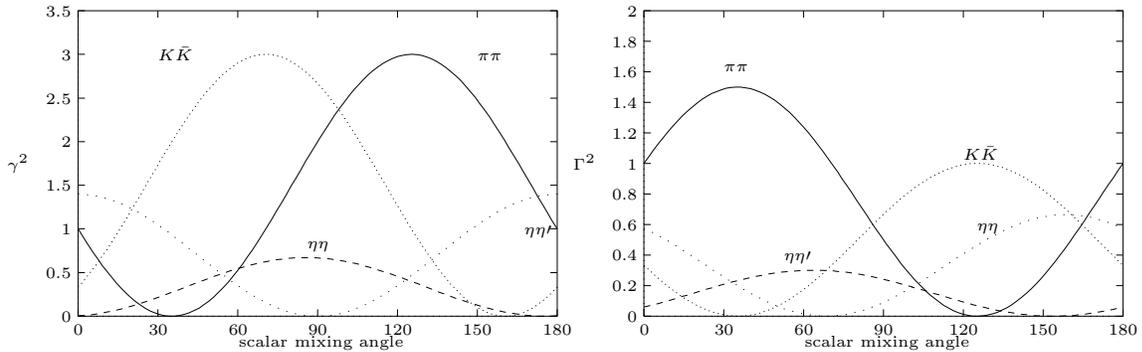

\begin{minipage}[t]{6.0cm}
\begin{picture}(0,0)%
\epsfbox{closefig.pstex}%
\end{picture}%
\setlength{\unitlength}{0.002000in}%
\begingroup\makeatletter\ifx\SetFigFont\undefined
\def\x#1#2#3#4#5#6#7\relax{\def\x{#1#2#3#4#5#6}}%
\expandafter\x\fmtname xxxxxx\relax \def\y{splain}%
\ifx\x\y   
\gdef\SetFigFont#1#2#3{%
  \ifnum #1<17\tiny\else \ifnum #1<20\small\else
  \ifnum #1<24\normalsize\else \ifnum #1<29\large\else
  \ifnum #1<34\Large\else \ifnum #1<41\LARGE\else
     \huge\fi\fi\fi\fi\fi\fi
  \csname #3\endcsname}%
\else
\gdef\SetFigFont#1#2#3{\begingroup
  \count@#1\relax \ifnum 25<\count@\count@25\fi
  \def\x{\endgroup\@setsize\SetFigFont{#2pt}}%
  \expandafter\x
    \csname \romannumeral\the\count@ pt\expandafter\endcsname
    \csname @\romannumeral\the\count@ pt\endcsname
  \csname #3\endcsname}%
\fi
\fi\endgroup
\begin{picture}(1444,899)(76,-132)
\put(247,-66){\makebox(0,0)[rb]{\smash{\SetFigFont{6}{7.2}{rm}0}}}
\put(247, 48){\makebox(0,0)[rb]{\smash{\SetFigFont{6}{7.2}{rm}0.5}}}
\put(247,163){\makebox(0,0)[rb]{\smash{\SetFigFont{6}{7.2}{rm}1}}}
\put(247,277){\makebox(0,0)[rb]{\smash{\SetFigFont{6}{7.2}{rm}1.5}}}
\put(247,391){\makebox(0,0)[rb]{\smash{\SetFigFont{6}{7.2}{rm}2}}}
\put(247,505){\makebox(0,0)[rb]{\smash{\SetFigFont{6}{7.2}{rm}2.5}}}
\put(247,620){\makebox(0,0)[rb]{\smash{\SetFigFont{6}{7.2}{rm}3}}}
\put(247,734){\makebox(0,0)[rb]{\smash{\SetFigFont{6}{7.2}{rm}3.5}}}
\put(266,-99){\makebox(0,0)[b]{\smash{\SetFigFont{6}{7.2}{rm}0}}}
\put(475,-99){\makebox(0,0)[b]{\smash{\SetFigFont{6}{7.2}{rm}30}}}
\put(684,-99){\makebox(0,0)[b]{\smash{\SetFigFont{6}{7.2}{rm}60}}}
\put(893,-99){\makebox(0,0)[b]{\smash{\SetFigFont{6}{7.2}{rm}90}}}
\put(1102,-99){\makebox(0,0)[b]{\smash{\SetFigFont{6}{7.2}{rm}120}}}
\put(1311,-99){\makebox(0,0)[b]{\smash{\SetFigFont{6}{7.2}{rm}150}}}
\put(1520,-99){\makebox(0,0)[b]{\smash{\SetFigFont{6}{7.2}{rm}180}}}
\put(109,334){\makebox(0,0)[b]{\smash{\SetFigFont{6}{7.2}{rm}$\gamma^2$}}}
\put(893,-132){\makebox(0,0)[b]{\smash{\SetFigFont{6}{7.2}{rm}
scalar mixing angle}}}
\put(1311,620){\makebox(0,0)[lb]{\smash{\SetFigFont{6}{7.2}{rm}$ \pi \pi$}}}
\put(475,620){\makebox(0,0)[lb]{\smash{\SetFigFont{6}{7.2}{rm}$ K \bar{K}$}}}
\put(893,128){\makebox(0,0)[b]{\smash{\SetFigFont{6}{7.2}{rm}$\eta \eta$}}}
\put(1416,163){\makebox(0,0)[lb]{\smash{\SetFigFont{6}{7.2}{rm}
$\eta \eta \prime$}}}
\end{picture}
\end{minipage}
\qquad \qquad
\begin{minipage}[b]{6.0cm}
\begin{picture}(0,0)%
\epsfbox{fnulzerfall.pstex}%
\end{picture}%
\setlength{\unitlength}{0.002000in}%
\begingroup\makeatletter\ifx\SetFigFont\undefined
\def\x#1#2#3#4#5#6#7\relax{\def\x{#1#2#3#4#5#6}}%
\expandafter\x\fmtname xxxxxx\relax \def\y{splain}%
\ifx\x\y   
\gdef\SetFigFont#1#2#3{%
  \ifnum #1<17\tiny\else \ifnum #1<20\small\else
  \ifnum #1<24\normalsize\else \ifnum #1<29\large\else
  \ifnum #1<34\Large\else \ifnum #1<41\LARGE\else
     \huge\fi\fi\fi\fi\fi\fi
  \csname #3\endcsname}%
\else
\gdef\SetFigFont#1#2#3{\begingroup
  \count@#1\relax \ifnum 25<\count@\count@25\fi
  \def\x{\endgroup\@setsize\SetFigFont{#2pt}}%
  \expandafter\x
    \csname \romannumeral\the\count@ pt\expandafter\endcsname
    \csname @\romannumeral\the\count@ pt\endcsname
  \csname #3\endcsname}%
\fi
\fi\endgroup
\begin{picture}(1444,899)(76,-132)
\put(247,-66){\makebox(0,0)[rb]{\smash{\SetFigFont{6}{7.2}{rm}0}}}
\put(247, 14){\makebox(0,0)[rb]{\smash{\SetFigFont{6}{7.2}{rm}0.2}}}
\put(247, 94){\makebox(0,0)[rb]{\smash{\SetFigFont{6}{7.2}{rm}0.4}}}
\put(247,174){\makebox(0,0)[rb]{\smash{\SetFigFont{6}{7.2}{rm}0.6}}}
\put(247,254){\makebox(0,0)[rb]{\smash{\SetFigFont{6}{7.2}{rm}0.8}}}
\put(247,334){\makebox(0,0)[rb]{\smash{\SetFigFont{6}{7.2}{rm}1}}}
\put(247,414){\makebox(0,0)[rb]{\smash{\SetFigFont{6}{7.2}{rm}1.2}}}
\put(247,494){\makebox(0,0)[rb]{\smash{\SetFigFont{6}{7.2}{rm}1.4}}}
\put(247,574){\makebox(0,0)[rb]{\smash{\SetFigFont{6}{7.2}{rm}1.6}}}
\put(247,654){\makebox(0,0)[rb]{\smash{\SetFigFont{6}{7.2}{rm}1.8}}}
\put(247,734){\makebox(0,0)[rb]{\smash{\SetFigFont{6}{7.2}{rm}2}}}
\put(266,-99){\makebox(0,0)[b]{\smash{\SetFigFont{6}{7.2}{rm}0}}}
\put(475,-99){\makebox(0,0)[b]{\smash{\SetFigFont{6}{7.2}{rm}30}}}
\put(684,-99){\makebox(0,0)[b]{\smash{\SetFigFont{6}{7.2}{rm}60}}}
\put(893,-99){\makebox(0,0)[b]{\smash{\SetFigFont{6}{7.2}{rm}90}}}
\put(1102,-99){\makebox(0,0)[b]{\smash{\SetFigFont{6}{7.2}{rm}120}}}
\put(1311,-99){\makebox(0,0)[b]{\smash{\SetFigFont{6}{7.2}{rm}150}}}
\put(1520,-99){\makebox(0,0)[b]{\smash{\SetFigFont{6}{7.2}{rm}180}}}
\put(109,334){\makebox(0,0)[b]{\smash{\SetFigFont{6}{7.2}{rm}$\Gamma^2$}}}
\put(893,-132){\makebox(0,0)[b]{\smash{\SetFigFont{6}{7.2}{rm}
scalar mixing angle}}}
\put(475,590){\makebox(0,0)[lb]{\smash{\SetFigFont{6}{7.2}{rm}$ \pi \pi$}}}
\put(1102,358){\makebox(0,0)[lb]{\smash{\SetFigFont{6}{7.2}{rm}$ K \bar{K}$}}}
\put(1137,174){\makebox(0,0)[lb]{\smash{\SetFigFont{6}{7.2}{rm}$\eta \eta$}}}
\put(614, 94){\makebox(0,0)[lb]{\smash{\SetFigFont{6}{7.2}{rm}
$\eta \eta \prime$}}}
\end{picture}
\end{minipage}
\vspace{1ex}
    \caption{
      Invariant couplings for the decay of the $f'_0$ meson into two
      pseudoscalar mesons: with the conventional OZI rule conserving
      flavor dependence (left), see also \protect\cite{amsclo}, and
      with the instanton induced three body vertex (right).  
    }\label{closedecay}
\end{figure}

The flavor dependent effective quark interaction used here was
computed by 't Hooft and others from instanton effects
\cite{Hoo76,SVZ80,Pet85}. 't Hooft showed that an expansion of
the (Euclidean) action around the one instanton solution of the
gauge fields with dominance of the zero modes of the fermion
fields leads to an effective interaction not covered by
perturbative gluon exchange. This interaction is chirally
symmetric, see also~\cite{now96}, but breaks the $\gamma_5$-
invariance and for three flavors it induces a six-point quark vertex
completely antisymmetric in flavor.  After normal ordering with
respect to the nontrivial QCD vacuum this leads to a
contribution to the constituent quarks masses, a two body
interaction
\begin{equation} \label{lag2}
 \Delta{\cal L}{(2)}(y) \propto
   g_{\mbox{eff}}^i      \,
    \varepsilon_{ikl} \, \varepsilon_{ik'l'} \,
    \overline{\Psi}(x) \, \overline{\Psi}(y)
            \bigl[ 1 \! \cdot \! 1 \,+\,
                   \gamma_5 \! \cdot \! \gamma_5
           \bigr]
       \left(2{\cal P}^C_{\bar{3}} + {\cal P}^C_6 \right)
       \Psi(x) \Psi(y) \Psi(y)
\end{equation}
and a three body term, see below. Here $i,k,l \in \{u,d,s\}$ are
flavor indices, and ${\cal P}^C$ are color projectors. This form
explicitly shows that this force only acts on antisymmetric
flavor states. The spin dependence is such, that this (contact)
force acts only for pseudoscalar and scalar mesons. The latter
contribution vanishes in the non-relativistic approximation we
used in a previous calculation~\cite{Bla90}, but in the present
relativistic model it leads to a sizeable flavor splitting in the
scalar spectrum, see Fig.~\ref{scalars}, similar to that observed
for the $\pi,\eta,\eta'$ pseudoscalars, see
Fig.~\ref{pseudoscalars}, but opposite in
sign~\cite{Mue94,Kle95}: In fact, the present model produces an
almost pure flavor singlet state $f_0^1$ at roughly 1 GeV
whereas the flavor octet states $f_0^8$, $a_0$, $K_0^*$ are
almost degenerate close to 1.5 GeV~, see Fig.~\ref{scalars}, for
a treatment in the Nambu, Jona-Lasinio model see~\cite{dmi96}.

Adopting this prediction then leads to the following
interpretation of experimental data as the $q\bar q$-flavor
nonet~\cite{Kle95}: We propose that the recently discovered
$f_0(1500)$, see also~\cite{koch96}, is not a glueball but the
scalar (mainly)--octet meson for which the $K\bar{K}$ decay mode
is suppressed as we will show below. The mainly--singlet state
could correspond to the broad $f_0(1000)$-state introduced by
Morgan and Pennington~\cite{Mor93a,Mor93b}, but there are also
arguments that it is to be identified with the $f_0(980)$. The
isovector and isodoublet states correspond to the $a_0(1450)$ and
$K^*_0(1430)$, respectively. Since the spectrum of scalar mesons
is rather puzzling~\cite{koch96} some comments are due: The
present calculation suggests, that one isoscalar and one
isovector scalar state at 1 GeV are not of the quarkonium type.
Indeed, coupled channel calculations performed by the J\"ulich
group~\cite{Jan95} (but see also~\cite{Wei90,toern}) suggest that
these resonances are related to $K\bar K$-dynamics. In this
spirit, the $f_0(1300)$ resonance is then the high energy part of
the broad $f_0(1000)$. For other resonances cited in the
literature the evidence is in general not so convincing and will
be not discussed here.

However, In particular the $f_0(1500)$~\cite{cb} was argued to
have properties incompatible with a pure $q\bar{q} $
configuration and was suggested to possess a large glue component
\cite{amslerclose,amsclo}. One of the major reasons for this
interpretation stems from the decay modes of the $f_0(1500)$ as
argued in \cite{amslerclose,amsclo}: There it is found to decay
into $\pi\pi \cite{gams},\; \eta\eta \cite{wa89},\;
\eta\eta^{\prime}$ \cite{toby} but not into $K\bar{K}$
\cite{gray}.  The $q\bar{q} $ hypothesis cannot fit these
branching ratios with a common $SU(3)_f$ scalar mixing angle,
when decaying through a conventional decay mechanism (see
Fig.~\ref{closedecay}) obeying Zweig's rule.  Furthermore, the full
width $\Gamma(f_0(1500)) =116\pm 17$~MeV seems to be incompatible
with a nonet structure: Taking the widths $\Gamma(a_0) = 270\pm
40$~MeV and $\Gamma(K_0^*) = 287\pm 23$~MeV as a scale for the
other members of the scalar nonet, a natural guess for the $f'_0$
width is around 500 MeV. The $f_0(1500)$ thus seems not naturally
to fit into the quarkonium nonet. In fact, we will argue, that
the same instanton induced forces can yield a decay pattern of
the $f_0(1500)$ with a strong $K \bar K$ suppression, without
assuming a glueball admixture. To this end we invoke the six
quark term from the instanton induced interaction which can be
written compactly with Weyl spinors $\Psi=(\xi,\eta)$, and spin
and color projection operators as (see~\cite{rit96} for details):
\begin{equation} \label{thooft}
 \Delta{\cal L}{(3)} \propto
   g_{\mbox{eff}}^{(3)} 
      \,\Bigl\{
    \,:\, \eta^{\dag} \, \eta^{\dag} \,\eta^{\dag} 
           \,{\cal P}^F_1 \,
           (2{\cal P}^{S}_{4} \otimes {\cal P}^{C}_{10}
           +5{\cal P}^{S}_{2} \otimes {\cal P}^{C}_{8})\,
       \xi \xi \xi\,:\,
      \Bigr\} + (\eta \longleftrightarrow \xi)\,.
   \nonumber 
\end{equation}
Upon calculating the lowest order contribution to the decay
amplitude of one meson into two mesons, on finds, that it only
acts if (pseudo)scalars are involved and, that the flavor
dependence of the instanton induced three-body interaction leads
to a selective violation of Zweig's rule: Only if a flavor
singlet participates, there is a contribution, of which the
flavor dependence deviates from that of the conventional decay
mechanism which obeys Zweig's rule. Actually this is quite
natural, since the substantial splitting and mixing of the
pseudoscalar states already indicates that in this sector Zweigs
rule must be violated. This implies, that the empirically very
successful OZI-rule, which was the basis of the argument to reject
the $f_0(1500)$ as a quarkonium state, can be selectively
circumvented in the decays of scalars into pseudoscalars. Indeed
Fig.~\ref{closedecay} shows quantitatively, that it is indeed
possible to account for the peculiar decay pattern of the scalar
states, if the SU(3)-mixing in the scalar nonet is small and
positive, which is indeed the case in our calculation. In
addition this new decay mechanism reduces also the calculated
width of $f_0^8$, although it is still too large when compared to
the experimental value~\cite{rit96}.

\section{Conclusion}\label{V}
In this contribution we presented the results of a covariant
constituent quark model, based on the Bethe--Salpeter equation,
and where confinement is implemented by a string like linear
potential explaining the Regge trajectories. An instanton induced
quark force explains not only the splitting and mixing of
pseudoscalar mesons, but suggest that such effects are also
present in the spectrum and a violation of the OZI rule in the
decays of scalar particles into pseudoscalars. We demonstrated
that a covariant treatment that takes into account the
relativistic components in the amplitudes is of utmost importance
when describing properties of deeply bound states and/or
processes at higher momentum transfer. At present we are applying
the same concepts in a covariant model of the three-quark system,
which is entirely possible but technically rather involved.

\section{Acknowledgment} \label{VI}
An important part of the material presented here was part of the
doctoral theses of C. M\"unz and J. Resag. We also acknowledge
contributions by W. Giersche, S. Hainzl and Ch. Ritter. We
highly appreciated the discussions with E. Klempt. Finally the
kind hospitality of Prof. L. Jarczyk and his crew in Cracow was
very much enjoyed.

\end{document}